\documentclass[conference]{IEEEtran}
\IEEEoverridecommandlockouts

\usepackage{cite}
\usepackage{amsmath,amssymb,amsfonts}
\usepackage{algorithmic}
\usepackage{graphicx}
\usepackage{textcomp}
\usepackage{xcolor}
\usepackage{orcidlink}
\usepackage[colorinlistoftodos,prependcaption,textsize=small]{todonotes}
\usepackage{amsmath}
\usepackage{multirow}
\usepackage{hhline}
\usepackage[most]{tcolorbox}
\usepackage{hyperref}
\usepackage{enumitem}
\setlist{nolistsep}

\newcommand{\approach}{\textbf{\texttt{AbsCon}}}
\newcommand{\llm}{LLM}
\newcommand{\llms}{LLMs}

\newcommand{\set}[1]{\mathcal{#1}}
\newcommand{\func}[1]{\mathtt{#1}}
\newcommand{\variable}[1]{\texttt{#1}}

\newcommand{\spec}{S}
\newcommand{\constraints}{\Phi}
\newcommand{\desc}{D}
\newcommand{\metamodel}{M}

\newcommand{\graph}{G}
\newcommand{\nodes}{\set{N}}
\newcommand{\edges}{\set{E}}
\newcommand{\labelFunc}{\func{L}}

\newcommand{\greedy}{\textbf{Direct}}
\newcommand{\mv}{\textbf{MV}}
\newcommand{\consistency}{\textbf{Con}}
\newcommand{\esc}{\textbf{ESC}}
\newcommand{\escf}{\textbf{ESC-F}}
\newcommand{\sr}{\textbf{SR}}
\newcommand{\fscore}{\textbf{F1}}
\newcommand{\accuracy}{\textbf{ACC}}
\newcommand{\precision}{\textbf{P}}
\newcommand{\recall}{\textbf{R}}

\def\BibTeX{{\rm B\kern-.05em{\sc i\kern-.025em b}\kern-.08em
    T\kern-.1667em\lower.7ex\hbox{E}\kern-.125emX}}
\begin{document}

\title{Accurate and Consistent Graph Model Generation from Text with Large Language Models
\thanks{\IEEEauthorrefmark{1}Work partially done during an internship at Huawei Research Canada.}
\thanks{We thank D\'aniel Varr\'o, Zohreh Aghababaeyan, Khaled Ahmed, Ru Ji, and anonymous reviewers for their valuable suggestions and feedback on the paper.}
}

\author{\IEEEauthorblockN{Boqi Chen\IEEEauthorrefmark{1}\IEEEauthorrefmark{2}\orcidlink{0000-0002-1451-3603}, Ou Wei\IEEEauthorrefmark{3}, Bingzhou Zheng\IEEEauthorrefmark{3}, Gunter Mussbacher\IEEEauthorrefmark{2}\orcidlink{0009-0006-8070-9184}}
\IEEEauthorblockA{\IEEEauthorrefmark{2}\textit{Electrical and Computer Engineering, McGill University, Canada} \\
\IEEEauthorrefmark{3}\textit{Huawei Research Canada, Canada} \\
boqi.chen@mail.mcgill.ca, \{ou.wei1,bingzhou.zheng\}@huawei.com, gunter.mussbacher@mcgill.ca	
}
}

\maketitle

\begin{abstract}
Graph model generation from natural language description is an important task with many applications in software engineering. With the rise of large language models (LLMs), there is a growing interest in using LLMs for graph model generation. Nevertheless, LLM-based graph model generation typically produces partially correct models that suffer from three main issues: (1) \textit{syntax violations}: the generated model may not adhere to the syntax defined by its metamodel, (2) \textit{constraint inconsistencies}: the structure of the model might not conform to some domain-specific constraints, and (3) \textit{inaccuracy}: due to the inherent uncertainty in LLMs, the models can include inaccurate, hallucinated elements. While the first issue is often addressed through techniques such as constraint decoding or filtering, the latter two remain largely unaddressed. Motivated by recent self-consistency approaches in LLMs, we propose a novel \textit{abstraction-concretization} framework that enhances the consistency and quality of generated graph models by considering multiple outputs from an LLM. Our approach first constructs a probabilistic partial model that aggregates all candidate outputs and then refines this partial model into the most appropriate concrete model that satisfies all constraints. We evaluate our framework on several popular open-source and closed-source LLMs using diverse datasets for model generation tasks. The results demonstrate that our approach significantly improves both the consistency and quality of the generated graph models. 


\end{abstract}


\section{Introduction}
\noindent
\textbf{Context.} Large language models (\llms{}), such as GPT-4~\cite{achiam2023gpt}, Llama3~\cite{dubey2024llama}, and DeepSeek-R1~\cite{guo2025deepseek}, have demonstrated impressive performance on various natural language tasks. Consequently, their application in model-driven engineering (MDE) has attracted significant interest~\cite{di2025use,camara2023assessment}. A common use case in MDE involves generating models from \emph{natural language descriptions}, for example, from requirements or use case narratives to domain models~\cite{chen2023automated,arulmohan2023extracting}, goal models~\cite{chen2023use}, sequence diagrams~\cite{ferrari2024model}, or taxonomies~\cite{chen2023prompting}. However, current approaches rely \emph{exclusively} on the \llm{} to generate or iteratively refine outputs, limiting their effectiveness. 


Traditional model generation aims to produce models that are \emph{consistent} with predefined well-formedness constraints~\cite{varro2018towards}. However, they do not account for natural language descriptions that allow domain experts to further specify the instance model. In contrast, \llm{}-based generation often prioritizes generating models that \emph{accurately} reflect such textual description, ignoring any constraints. Although \emph{syntactical} correctness is typically addressed in both cases, ensuring both accuracy and consistency is critical for effectively utilizing generated models in subsequent model-driven automation processes, such as code generation~\cite{garzon2015umple} and model-based testing~\cite{gurbuz2018model}. Therefore, jointly considering both the accuracy and consistency of models generated by LLMs is essential.

\noindent
\textbf{Problem statement.}
Due to the intrinsic limitations of \llms{}, the models they generate are typically \emph{partially correct} and may exhibit issues in all three aspects: (1) \textbf{syntax}: the textual representation of the model may violate the syntax specified by the metamodel; (2) \textbf{consistency}: the structure of the graph model may conflict with domain-specific well-formedness constraints~\cite{varro2018towards}; and (3) \textbf{quality}: the generated graph models may be inaccurate with respect to the given descriptions due to hallucinations from \llms{}~\cite{chen2023automated,chen2023use}. Taking UML activity diagrams as an example, the first issue occurs when generated models violate modeling language syntax (e.g., PlantUML~\cite{plantuml}). Models affected by the second issue may contain errors like a decision node with only one outgoing edge, which is not valid in activity diagrams. Finally, the third issue occurs when an activity diagram includes inaccurate information compared to the description such as incorrect ordering of activities. The first issue can typically be mitigated by incorporating a filtering mechanism for the outputs or by applying a \emph{constraint decoding} method to enforce syntax rules~\cite{netz2024using,mousavi2024towards}. However, approaches to address the second and third issues have not yet been thoroughly investigated. 

Self-consistency leverages the inherent uncertainty of \llms{} by synthesizing a more accurate result from multiple outputs generated with the same input~\cite{wang2023self}. This approach improves the performance of \llms{} on a variety of tasks, particularly when combined with the chain-of-thought (CoT) approach~\cite{wei2022chain}. With foundational CoT-based reasoning \llms{}, such as OpenAI o1~\cite{openai2024learning} and DeepSeek R1~\cite{guo2025deepseek}, the adoption of self-consistency is expected to become more common across various applications. The core hypothesis of self-consistency is that if an \llm{} is capable of solving a task, it should produce the correct answer more frequently than incorrect alternatives~\cite{wang2023self}. However, this assumption does not necessarily hold for model generation. Since graph models are composite structures, even if an \llm{} has the potential to solve a task, it may only predominantly output \emph{partially correct graphs}.

\noindent
\textbf{Objective.}
Our paper aims to simultaneously address the consistency and quality issues when applying \llms{} for graph model generation. 
Motivated by \llm{} self-consistency, we combine multiple \llm{}-generated models given a problem description, a metamodel, and a set of well-formedness constraints, such that the final graph adheres to the metamodel and constraints while accurately reflecting the input description. 

We assume that if \llms{} are capable of solving a model generation task, they will produce the \emph{correct graph elements} more frequently. By aggregating multiple outputs and applying constraints on the graph structure, our approach derives improved solutions. More generally, we aim to demonstrate that leveraging the intrinsic uncertainty of \llms{} by combining their generative capability with formal constraints in modeling techniques can enhance both the \textbf{consistency} (issue 2) and \textbf{quality} (issue 3) of the generated models.

\noindent
\textbf{Contribution.}
We present \approach{}: an \textbf{\texttt{Abs}}traction-\textbf{\texttt{Con}}cretization framework motivated by self-consistency to derive an accurate and consistent graph from multiple generation results based on partial modeling. Multiple \llm{}-generated graphs are first \emph{abstracted} into a probabilistic partial model and then \emph{concretized} into a final model taking the metamodel and constraints into consideration. Specifically, this paper makes the following contributions:
\begin{enumerate}
    \item We propose \approach{}: a general framework to create an accurate and consistent model from a pool of candidates.
    \item We propose a specific abstraction and concretization method that guarantees the consistency of generated models while improving their quality.
    \item We apply \approach{} to three different model generation tasks at different levels of complexity and domains.  
    \item Through systematic evaluation, we demonstrate the effectiveness of \approach{} in generating consistent models while also improving model quality. Moreover, we explore the influence of the number of candidates on the final model and provide insights on further improvements. 
\end{enumerate}

\textbf{Added value.}
The \approach{} approach guarantees consistency in the generated models, which in turn enhances the overall quality of the outputs. Since it treats LLMs as a black box and requires no additional training or modification to their structure, \approach{} can be easily applied to various model generation tasks. \approach{} can be considered as a novel approach to improve LLMs' performance in model generation through test-time compute~\cite{snell2024scaling}, which is flexible and can be further refined through different approaches at each stage.









\section{Background}
\subsection{Graph generation}
\label{sec:graph_generation}
Graph generation is a well-studied task in various software engineering fields, including model-driven engineering (MDE)~\cite{saini2022machine,semerath2021automated,varro2018towards} and machine learning \cite{zhu2022survey}. It aims to generate a graph from a given specification while adhering to predefined structures, often expressed as constraints. 

\noindent
\textbf{Graphs.}
For simplicity, this paper ignores node attributes and defines a \emph{labeled graph}. A labeled graph is represented as a tuple $\graph{}=(\nodes{}, \edges{}, \labelFunc{})$ where $\nodes$ is the set of nodes, $\edges: \nodes{} \times \nodes{}$ is the set of directed edges, and the mapping $\labelFunc{}: \nodes{} \cup \edges{} \to \set{T}$ assigns a textual label to each node and edge. 

\noindent
\textbf{Uncertainty in graphs.} 
Uncertainty in graphs is often captured using partial graph models \cite{famelis2012partial,salay2012language,famelis2011partial,marussy2020specification}. This uncertainty typically arises during intermediate stages of model generation \cite{semerath2018graph}. In a partial model, each element is associated with a three-valued logic: $1$ indicates that the element must appear in the concrete model, $0$ means that it must not appear, and $\frac{1}{2}$ denotes that the element \emph{may} be included in the final model.


\noindent
\textbf{Graph generation.}
Graph generation is the task of producing a graph that best conforms to a given specification $\spec{}$. Typically, the specification consists of a metamodel $\metamodel{}$ and a set of constraints $\constraints{}$. For text-based model generation, a natural language description $\desc{}$ is also included: $\spec{}=(\metamodel{}, \constraints{}, \desc{})$.


\subsection{Self-consistency for large language models}
\label{sec:selfcon}
Self-consistency \cite{wang2023self} improves LLM performance by generating multiple answers and then selecting the final answer via majority voting. Formally, let $\func{llm}$ be an LLM, $p$ the prompt template, and $t$ the task input. Self-consistency assumes that the final answer is chosen from a fixed set of candidates. Suppose there are $m$ candidate outputs $a_i \in \set{A}, i \in [1,m]$. Since these answers are usually generated along with reasoning paths via chain-of-thought \cite{wei2022chain}, each answer $a_i$ is paired with a corresponding reasoning path $r_i$. Given the inherent uncertainty of LLM outputs, the output of an LLM for a given prompt and task can be modeled as a probability distribution over both reasoning paths and final answers: $\func{llm}(p, t) \sim P(r_i, a_i | p, t)$. Self-consistency estimates the \emph{marginalized probability distribution} $P(a_i \mid p, t)$ by sampling multiple $(r_i, a_i)$ pairs and selecting the most probable answer using a voting mechanism. Specifically, majority voting, where each candidate is weighted equally, outperforms alternative voting schemes \cite{wang2023self}.

However, this method typically assumes simple outputs such as classification labels or numbers, making them unsuitable for complex outputs like graph models. While extensions of self-consistency for more general LLM outputs have been proposed~\cite{chen2023universal,thirukovalluru2024atomic}, these approaches solely rely on LLMs. Hence, they are unsuitable for model generation tasks where consistency with the constraints is critical. 

In this work, we propose a novel self-consistency-based method for model generation based on majority voting. 


\subsection{Constraint optimization}
\emph{Constraint optimization} involves finding values for a set of variables that optimize an objective function while satisfying \emph{all constraints}. Let $\set{V} = \{v_1,...,v_n\}$ be a set of variables, $\mathcal{C}$ the constraints over these variables, and $\func{Obj}$ an objective function. The goal is to determine an assignment for $\set{V}$ that maximizes $\func{Obj}$ while satisfying $\mathcal{C}$: $\operatorname*{argmax}_{\set{V}} \func{Obj}(\set{V}), s.t. \set{V} \models \mathcal{C}$.


Many solvers are available for this type of optimization problem~\cite{forrest2024cbc,holmstrom2009user,gurobi2024}. 
Well-formedness constraints can often be expressed as logical constraints compatible with these solvers.
In this paper, we focus on \emph{linear constraints} since they cover a wide range of practical scenarios and are efficient to solve. Nonetheless, our framework can be adapted with \emph{any} type of constraints and corresponding solvers.

\section{Approach}

\noindent
\textbf{Problem description.}
We address the problem of (instance) graph model generation from a specification $S = (M, \Phi, D)$ with a metamodel $M$, a constraint set $\Phi$, and a textual description of an instance model $D$. Let $G^{gt}$ be a ground truth model that is consistent with the metamodel and constraints while compliant with the description: $G^{gt}\models (M, \Phi) \land G^* \sim D$. The goal of model generation is to identify a generator $\texttt{Gen}$ that generates a model $G^*$ approximating the ground truth $\texttt{Gen}(S)=G^*\approx G^{gt}$. To tackle this challenge, this paper proposes an approach leveraging a set of output models from LLMs. Let $\mathcal{G}=\{G_1, \ldots, G_n\}$ be a set of $n$ candidate models generated by an \llm{}, each potentially covering some aspects of the description $D$ but violating some constraints in $M$ or $\Phi$. Our approach identifies a final graph model $G^*$ that best aligns with the candidate set $\mathcal{G}$ while satisfying $M$ and all constraints in $\Phi$. Note that each candidate graph output by an \llm{} is represented in a textual language (e.g., PlantUML \cite{plantuml} or Mermaid \cite{knut2024mermaid}).


Unlike the instance model generation addressed in this paper, traditional model generation tasks create models from a metamodel and constraints~\cite{varro2018towards}, without using natural language descriptions. As a result of this conceptual limitation, traditional methods~\cite{soltana2020practical,semerath2021automated,semerath2018graph,marussy2024refinery} are not directly applicable in this context.



\subsection{Motivating example}
\begin{figure}
    \centering
    \includegraphics[width=0.85\linewidth]{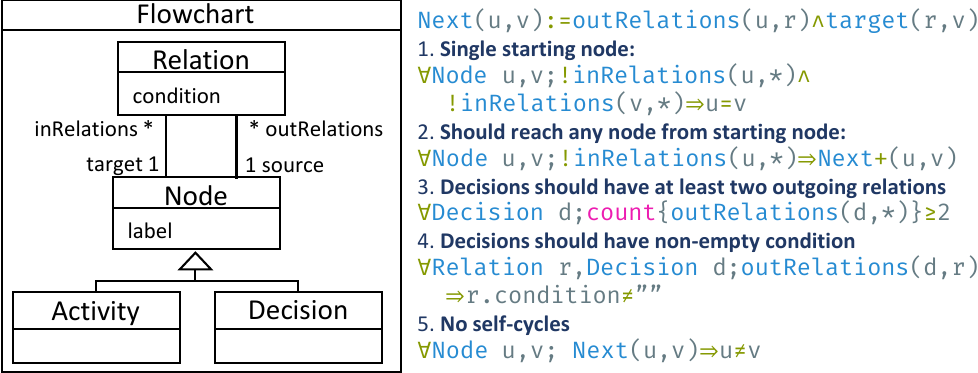}
    \caption{The metamodel and constraints for flowcharts}
    \label{fig:running_example}
\end{figure}

\begin{figure*}
    \centering
    \includegraphics[width=0.9\linewidth]{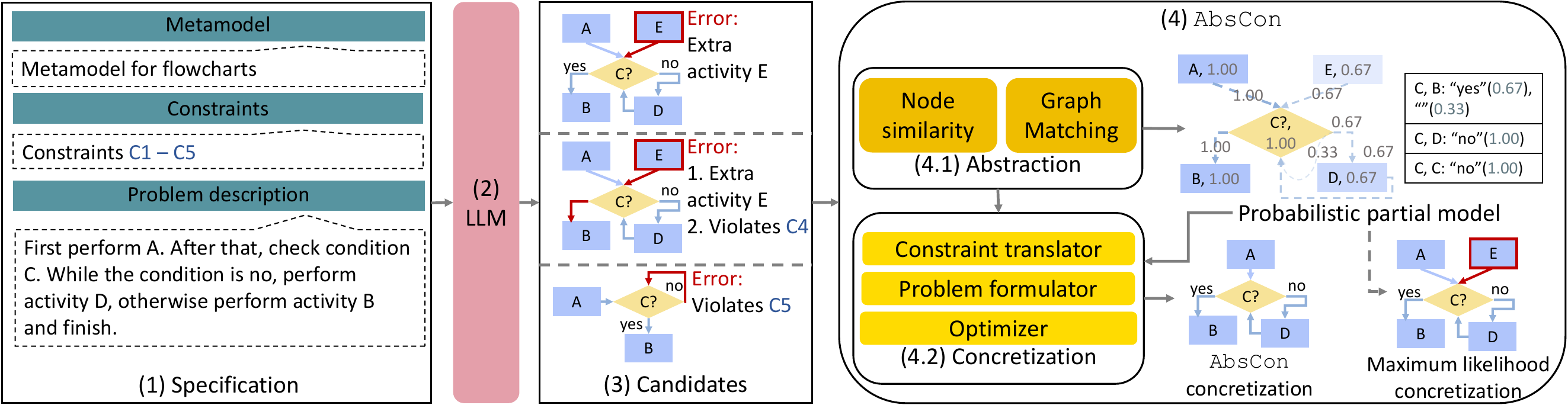}
    \caption{Overview of the \approach{} approach with a flowchart generation example}
    \label{fig:architecture}
\end{figure*}

\autoref{fig:running_example} illustrates the setup for constructing flowcharts, including both the metamodel and the well-formedness constraints defined in a simplified specification language \cite{marussy2020specification}. The metamodel comprises two types of nodes: \emph{Activity} and \emph{Decision}. Each node has a single attribute, \emph{label}, representing the activity content or decision criteria. Nodes are connected via \emph{Relations}, with outgoing relations from decision nodes optionally including a condition. Parallel flows are represented by activities with multiple outgoing connections. 

This domain includes five constraints, which define the \emph{semantic validity} of an instance flowchart. Specifically, a valid instance flowchart must: (1) have a single starting node; (2) allow reaching every other node from the starting node; (3) require decision nodes to have at least two targets; (4) ensure that each outgoing relation from a decision node has a non-empty condition; and (5) contain no self-cycles. 

\autoref{fig:architecture}.(1)-(2) illustrates the typical setup for LLM-based model generation. In this context, the input prompt includes the metamodel of the flowchart, its constraints, and the \emph{problem description}, while the LLM is asked to generate a flowchart that adheres to these specifications. 

Due to the uncertainty of \llms{}, all generated models are only \emph{partially correct} (candidates 1-3). For example, the LLM may hallucinate a non-existent activity E (candidates 1 and 2) or produce models that violate some constraints (candidates 2 and 3). However, each \llm{}-generated model often captures different correct aspects of the description. Consequently, a better overall solution can be obtained through appropriate combination of these solutions. This paper focuses on synthesizing such a combination.


\subsection{Overview}

\autoref{fig:architecture} provides an overview of \approach{}. The input to the \llm{} consists of a \emph{specification} (\autoref{fig:architecture}.(1)), accompanied by few-shot examples or CoT instructions. The \textit{\llm{}} (\autoref{fig:architecture}.(2)) then generates a set of \emph{candidate graphs} (\autoref{fig:architecture}.(3)). To leverage the partially correct aspects of these candidates, \approach{} first constructs a \emph{probabilistic partial model} that integrates all candidate graphs (\autoref{fig:architecture}.(4.1)). This partial model is subsequently \emph{concretized} into a final output by searching for the optimal model ((\autoref{fig:architecture}.(4.2)). 

\subsection{Candidate generation}
\noindent
\textbf{Prompt.}
The input \emph{prompt} provided to the LLM comprises a specification, including a metamodel, well-formedness constraints, and description of the model. Optionally, the prompt may include additional information created using various prompting techniques, such as few-shot examples or CoT instructions. For example, in \autoref{fig:architecture}.(1), the problem description is a paragraph detailing the behavior of the system, while the output graph is required to be a valid flowchart. 


\noindent
\textbf{Candidates.}
To identify promising sub-graphs in LLM outputs using self-consistency, multiple candidates are generated. Due to the benefits described in \autoref{sec:selfcon}, we adapt the original self-consistency setup, which produces multiple outputs from the same input prompt using a non-zero temperature \cite{wang2023self}. Nevertheless, \approach{} is independent of the candidate generation method. 

In \approach{}, candidates need to be parsed as graphs, which requires that the generated models adhere to the metamodel (i.e., contain no syntactical errors). In early experiments, we observe that LLMs rarely produce syntax errors when using the Mermaid diagramming language, a widely adopted tool for visualizing various types of models \cite{knut2024mermaid}. Therefore, we use Mermaid as the output language and filter out any generated models with syntax errors. Alternatively, one may use constraint decoding \cite{netz2024using} to ensure syntactical correctness.  

Three example candidate flowcharts are shown in \autoref{fig:architecture}.(3). Although each candidate contains \emph{correct sub-graphs}, none of them is fully correct. The goal of \approach{} is to construct an improved final output by considering these correct sub-graphs.



\subsection{Abstraction}
\noindent
\textbf{Element similarity.}
The abstraction module (\autoref{fig:architecture}.(4.1)) integrates all candidate elements to construct a \emph{partial model} covering all candidate models. To merge elements from two models, we first calculate the similarity between their corresponding graph elements. Similarity can be measured using techniques such as node attribute embedding \cite{chen2024embedding} or string edit distance. Since string edit distance focuses only on character-level similarity, we use pre-trained text embeddings to better capture semantic similarity between node labels.

We assume that each node in the model includes an identification attribute, such as a label, which is common in many models (e.g., class names in domain models, activity names in activity diagrams, or concept names in taxonomies). Even if duplicate labels exist, the embedding similarity between two node labels still provides a useful \emph{hint} for matching nodes across models. In contrast, since relations may have empty labels, we use an \emph{exact match} approach for assessing relation similarity: two relations are considered a match if their labels and both their source and target nodes match exactly. 


\noindent
\textbf{Graph matching.}
Element-wise similarity provides an initial local indication for matching corresponding components between two models. The structure of the graph still needs to be considered to capture the global similarity. We formulate this graph matching program as a \emph{graph edit distance} problem, where the distance between matched elements is determined by their element-wise similarity. A graph edit path specifies the operations such as matching, addition, or deletion, required to transform one graph into another, naturally yielding a mapping between their elements. This problem can be addressed using existing optimized methods \cite{abu2015exact}. Although the problem is NP-hard for arbitrary graphs, in practice, near-optimal matches are typically identified in a short amount of time. 

\noindent
\textbf{Probabilistic partial model.}
Traditional partial models represent uncertainty using three-valued logic (see \autoref{sec:graph_generation}). However, this formulation does not capture the \emph{likelihood} of each element, which is a crucial aspect for determining the frequency of model elements. We adapt the classic \emph{partial model} with \emph{probabilities} that indicates the likelihood of each element's existence in the concretized model. We refer to this extended model as \emph{probabilistic partial model}. 

Formally, a probabilistic partial model is defined as a tuple $\mathbb{\graph{}}=(\nodes{}, \edges{}, \func{L}, \func{P})$, where $\func{L}: \nodes{} \cup \edges{} \to \mathtt{T}$ is the \emph{probabilistic label mapping} that assigns to each node and edge a probability distribution over labels $\mathcal{T}$. That is, for any element $e \in \nodes{} \cup \edges{}$, $\func{L}(e)$ is a function $\mathtt{T}_e: \set{T}_e \to [0, 1]$ satisfying $\sum_{t \in \set{T}_e} \mathtt{T}_e(t)=1$. $\func{P}: \nodes{} \cup \edges{} \to [0, 1]$ is a mapping that assigns the likelihood of \emph{existence} for each element.

\textbf{Partial model construction.}
Given a set of candidate models, we propose an \emph{incremental} method to build the partial model using graph matching. We begin by selecting one candidate as the \emph{initial partial model}, initializing each of its elements with a count of 1. The remaining candidates are then matched and merged into this seed model. For a node $n$ from a candidate model and a node $m$ from the partial model, there are two possible scenarios: 
\begin{enumerate}
    \item If $n$ matches a node $m$ in the partial model, increment $m$'s count by 1 and add $n$'s label to $m$'s list of possible labels. Update the representative label of $m$ to the \emph{most frequent label} in the list. 
    \item If $n$ does not match any node in the partial model, add $n$ to the partial model with an initial count of 1.
\end{enumerate}
Furthermore, if a node $m$ in the partial model has no corresponding match in the candidate model, no action is needed. The same procedure also applies to relation updates. 

The \emph{probability} of an element is calculated by dividing its count by the total number of candidate models. When an element is associated with multiple labels, the probability for each label is determined by dividing the number of occurrences of that label by the element's total count. Note that alternative abstraction approaches, such as Bayesian techniques, can also be used to construct the partial model. We leave this exploration as future work. The output of \autoref{fig:architecture}.(4.1) shows the constructed partial model for the three candidate flowcharts.

\subsection{Concretization}
\noindent
\textbf{Graph consistency.}
The probabilistic partial model captures the likelihood of each element's existence across candidate models. A naive concretization approach would apply majority voting that selects elements appearing in most candidates. However, this method ignores the constraints over the graph structure. For example, as shown in the maximum likelihood concretization of \autoref{fig:architecture}, majority voting erroneously adds node E, which violates the consistency constraint (C1: single source) and fails to remove the hallucinated node. To overcome these issues, we propose a \emph{constraint-aware} concretization method (\autoref{fig:architecture}.(4.2)). Specifically, we formulate the final output selection as a \emph{constraint optimization} problem that can be efficiently solved using existing solvers.

\noindent
\textbf{Constraint translator.}
In this paper, we manually translate the metamodel and well-formedness constraints into first-order logic (FOL) formulae, which can be processed by many existing optimization solvers. Typically, such FOL formulae can be automatically derived from high-level graph constraint languages like Object Constraint Language or VIATRA Query Language \cite{bergmann2014translating,semerath2017formal} and have been previously used to encode constraints in neural network frameworks \cite{chen2022consistent}.


\noindent
\textbf{Problem formulator.}
The problem formulator defines the optimization problem using the partial model, metamodel, and constraints. 
Given a partial model $\mathbb{\graph{}}=(\nodes{}, \edges{}, \func{L}, \func{P})$, we define a set of \emph{decision variables} as $\set{X} = \{\variable{x}_e, \forall e \in \set{E}\} \cup \{\variable{x}_n, \forall n \in \set{N}\}$, where $\variable{x} = 1$ indicates that the corresponding element is included in the final model and $\variable{x} = 0$ otherwise.

The constraints $\constraints{}$ expressed as first-order logic (FOL) formulae are applied to these decision variables, yielding a set of logical formulae $\constraints{}'$. In addition, we introduce extra constraints to link nodes and relations. Specifically, for each relation $e=(s, t) \in \set{E}$, we require that if either the source node $s$ or the target node $t$ is not selected, then $e$ must be excluded. Thus, the overall set of constraints is given by $\mathcal{C} = \constraints{}' \cup \{e=(s, t) \in \set{E}| \variable{x}_s=0 \lor \variable{x}_t = 0 \implies \variable{x}_e = 0\}$.

Under the Naive Bayes assumption, where the probabilities of individual elements are independent and the inter-element relations are expressed as constraints, we use the \emph{binary cross-entropy} as the optimization objective. While other objectives exist, cross-entropy balances simplicity and effectiveness. The constraint optimization problem is thus formulated as:  
\begin{equation*}
\scalebox{0.85}{
$\operatorname*{max}_{\set{X}} \sum_{a \in \set{N}\cup\set{E}}\variable{x}_a \log{\func{P}(a)} + (1 - \variable{x}_a)\log{(1 - \func{P}(a))}, s.t.\: \set{X} \models \mathcal{C}$
}
\end{equation*}
 
The solution to this problem represents the \emph{optimal} concretization of the partial model while satisfying all constraints. Failure to obtain a feasible solution implies that \emph{no combination} of candidates can produce a consistent graph, which may indicate that the LLM is not capable of this task or that more candidates are needed. As shown in the output of \autoref{fig:architecture}.(4.2), the final model obtained using \approach{} successfully avoids including the hallucinated activity E by optimizing for maximum probability under the given constraints. 






\section{Case Studies}
Flowcharts serve as example behavioral models targeted by model generation. In addition to this use case, we evaluate the effectiveness of \approach{} on two other model types: structural and executable. In this section, we briefly describe these cases. 

\begin{figure*}
    \centering
    \includegraphics[width=0.9\linewidth]{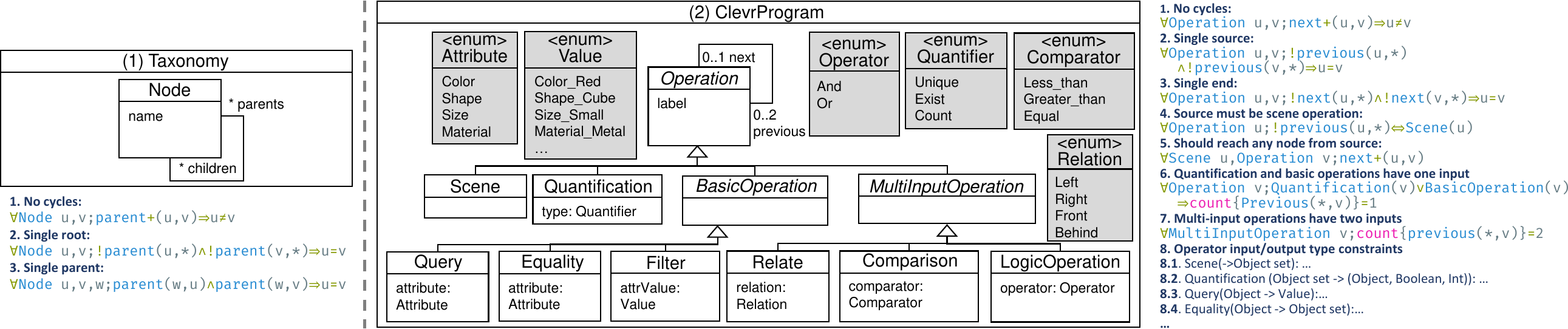}
    \caption{Metamodels and constraints for taxonomies and executable program graphs}
    \label{fig:case_studies}
\end{figure*}

\subsection{Taxonomy generation}
Taxonomy generation is a classical modeling task that creates a hierarchical structure from a given set of concepts. Structural constraints, however, may vary with the domain~\cite{cocos2018comparing}. In this task, the set of node labels is predefined. 

\autoref{fig:case_studies}.1 shows the metamodel and constraints for taxonomies. The metamodel specifies that concepts are connected through parental relations. Consequently, three constraints are defined: (1) the taxonomy must be acyclic, (2) it must contain only one root concept (i.e., a concept without any parent), and (3) each non-root concept only has a single parent.

\subsection{Program induction}
Program induction is the task of converting a description into an \emph{executable} program graph. This paper focuses on program graphs in the Clevr dataset \cite{johnson2017clevr}. Originally designed for visual question answering, the Clevr dataset contains images, scene graphs, and questions. Each object contains four attributes: color, shape, size, and material with various possible values. Moreover, objects are related by spatial relations such as \emph{left} and \emph{behind}. Questions are translated into program graphs that can be executed on a scene to derive an answer.

The metamodel and constraints for the Clevr program are presented in \autoref{fig:case_studies}.2. A program consists of a sequence of interconnected operations. Each operation either queries an object's attribute or retrieves a set of objects based on a specific relation. For the graph to be executable, it must satisfy a set of constraints. Specifically, (1) the graph must be acyclic with (2) a single entry point and (3) a single exit point. Additionally, the program must (4) begin with the \emph{Scene} operation to retrieve the input scene and (5) ensure that all operations are reachable. Operations also (6-7) require a varying number of inputs depending on their type, and an input mapping is defined so that each operation's input types match the output types of its preceding operation. Moreover, Operators also (8) require specific input and output types (not shown in the figure for brevity). For example, a \emph{query} operation must receive an object and produces an attribute value, whereas \emph{quantification} accepts a set of objects and produces an object when the quantifier is `unique', a boolean value for `exist', or an integer for `count'.

\section{Evaluation}

\subsection{Research questions}
In the evaluation section, we assess the effectiveness of \approach{} by addressing the following research questions:

\begin{enumerate} 
    \item How do the consistency and quality of \approach{}'s outputs compare to those of alternative approaches? 
    \item How does the consistency of \llm{}-generated models impact model quality? 
    \item How does the number of candidates affect \approach{}'s performance? 
\end{enumerate}

\subsection{Evaluation setup}
\paragraph{Target datasets}
We evaluate the effectiveness of our approach using the following datasets of the case studies.

\noindent\textbf{Flowcharts.}
We derive a set of flowcharts from the PAGED dataset \cite{du2024paged}, a high-quality, automatically generated collection of procedure graphs based on input descriptions. These graphs extend traditional flowcharts by incorporating data flow and multiple actors. We only select flowcharts from the PAGED dataset for this experiment. 

A portion of the dataset consists of trivial graphs without branches or forks, which are removed during evaluation. After filtering, we obtain 331 non-trivial description-graph pairs.
These procedure graphs are then converted into Mermaid flowcharts. 
Finally, we randomly select five cases as few-shot examples and use the rest for evaluation.

\noindent\textbf{Taxonomies.}
We use the WordNet taxonomy dataset \cite{miller1995wordnet} for the taxonomy construction task. WordNet is a hypernym dataset consisting of general English terms, forming 761 taxonomies, each containing between 11 and 50 terms. We randomly sample 100 taxonomies for evaluation and select three as few-shot examples.

\noindent\textbf{Program graphs.}
To assess the effectiveness of our approach for \textit{executable} models, we use questions and scenes from Clevr \cite{johnson2017clevr}. In this benchmark, an input question is translated into an executable graph, which is then run on a scene to generate an answer. The questions fall into three categories: (1) count questions, where the answer is a number; (2) judge questions, with binary answers; and (3) query questions, where the answer is an object attribute value. We randomly select 100 questions for each category, resulting in 300 questions for evaluation, and designate four as few-shot examples.

\paragraph{LLMs}
We evaluate our approach using four popular LLMs with varying capabilities. Two of these models, GPT-4o-mini \cite{gpt-4o-mini} and GPT-4o \cite{hurst2024gpt}, are the latest in OpenAI's GPT family, demonstrating enhanced performance on complex tasks requiring advanced reasoning. However, as these models are closed-source and accessible only via APIs, they may not be suitable for scenarios involving sensitive data. To address this, we also evaluate our approach using two variants of the open-source \llm{} Llama3.1 (8B and 70B) \cite{dubey2024llama}.

For model generation, we adopt few-shot CoT~\cite{wei2022chain}. We include few-shot examples with manually crafted chain-of-thought reasoning steps embedded in each prompt. These examples, along with the necessary specifications, are provided as input to the LLM to generate candidate models.

\paragraph{Compared methods}
We compare \approach{} with several alternative methods. The direct CoT approach (\greedy{}) approximates greedy decoding, where the token with the highest probability is selected at each step. However, since determinism cannot be guaranteed for OpenAI models, we set the temperature parameter to a very low value (i.e., $0.01$).

The self-consistency with the majority voting approach (\mv{}) applies the original self-consistency method \cite{wang2023self} to each graph element, similar to atomic self-consistency \cite{thirukovalluru2024atomic}, by performing majority voting on the relations (including the source and target nodes) within the graph models.

For executable models in Clevr, self-consistency can also be applied to the \emph{execution output}, a method known as \emph{execution-based self-consistency} (\esc) \cite{li2022competition,shi2022natural}. However, as generated graphs may violate constraints and cause execution errors, we introduce a stronger baseline that integrates execution-based self-consistency with a filtering mechanism (\escf). In this approach, graphs that result in execution errors are excluded from the majority voting. If all candidates are inconsistent, the method will output an error. 


\paragraph{Metrics}
\label{sec:metrics}
The generated graph models are evaluated based on two criteria: (1) \textbf{consistency}, which measures the percentage of models that are \emph{fully} compliant with all constraints, and (2) \textbf{quality}, which assesses how accurately the generated graphs match the given description. Since automated comparison of a graph model with the natural language description may be unreliable, we evaluate model quality using either downstream task performance or ground truth comparison, depending on the use case.

For behavioral models in the PAGED dataset and structural models in WordNet, the \textit{consistency ratio} (\consistency) is defined as the percentage of graphs that satisfy all constraints. For executable models in the Clevr benchmark, we use the \textit{success rate} (\sr) as the consistency metric, representing the percentage of graphs that can be successfully executed without error.

Evaluating the quality of non-executable graph models is challenging due to their complexity and potential label ambiguities. To address this, we use the ground truth models in the datasets as reference models and compare the relation sets extracted from the generated and reference graphs. Each relation is represented by the labels of the source node, the target node, and, if applicable, the relation label. To account for possible label ambiguities, we use \emph{soft precision, soft recall}, and \emph{soft F1-scores} \cite{franti2023soft}. These metrics extend standard evaluation by redefining set calculations to allow soft comparisons between elements instead of strict matching. Given a set of predicted relations $\set{E}$ and a set of reference relations $\set{E}_r$, the \emph{soft cardinality} of a relation set $\set{E}$ is defined as:
\begin{equation*}
\scalebox{0.85}{
$ \variable{card}(\set{E}) = \sum_{e \in \set{E}} \frac{1}{\sum_{e' \in \set{E}} \variable{Sim}(e, e')}$
}
\end{equation*}
where $\variable{Sim}$ is some similarity measure. The cardinality of $\set{E} \cap \set{E}_r$ can be defined as $\variable{card}(\set{E} \cap \set{E}_r) = \variable{card}(\set{E}) + \variable{card}(\set{E}_r) - \variable{card}(\set{E} \cup \set{E}_r)$, where $\set{E} \cup \set{E}_r$ represents the union of elements from both sets. Using soft cardinality, the soft precision (\precision), soft recall (\recall), and soft F1-score (\fscore{}) are computed using the standard definitions with soft cardinality:
\begin{equation*}
\scalebox{0.8}{
$\precision{}= \frac{\variable{card}(\set{E} \cap \set{E}_r)}{\variable{card}(\set{E})}, \recall{} = \frac{\variable{card}(\set{E} \cap \set{E}_r)}{\variable{card}(\set{E}_r)}, \fscore{} = \frac{2 \precision{} \cdot \recall{}}{\precision + \recall}$
}
\end{equation*}

When the similarity measure $\variable{Sim}$ is exact match, the metrics reduce to their standard definitions. For taxonomy construction, where ground truth node labels are provided, we use exact match as the similarity measure. For flowcharts, we adopt \emph{token overlap} as the similarity measure. Let $\set{T}_{e1}$ and $\set{T}_{e2}$ denote the sets of tokens associated with relations $e1$ and $e2$, respectively. The token overlap similarity is defined as:
\begin{equation*}
\scalebox{0.8}{
$\variable{Sim}_{token}(e1, e2) = \frac{\set{T}_{e1} \cap \set{T}_{e2}}{\set{T}_{e1} \cup \set{T}_{e2}}$
}
\end{equation*}
We do not use embedding similarity since it is used during the abstraction step to avoid potential bias. For executable models, we measure accuracy (\accuracy{}) of the answer from execution. 

\paragraph{Implementation details}
In \approach{}, we use the \emph{all-MiniLM-L6-v2} encoder from Sentence Transformers \cite{reimers2019sentence} to encode labels for element similarity. During concretization, the constraint optimization problem is solved using the CBC solver \cite{forrest2024cbc}. For token overlap measurements, tokens are generated using the GPT-4o tokenizer. Additionally, when generating each candidate, we set the LLM temperature to 0.7 and enforce a 5-second timeout for computing graph edit distance. The paper artifacts, including the prompt used in the experiments, are available at \cite{artifact_models2025}.

\subsection{RQ1: Graph quality and consistency}

\begin{table*}[tb]
    \centering
    \caption{Performance of compared approaches on different types of graph models (in \%)}
    \label{tab:rq1}
    \scalebox{0.85}{
    \begin{tabular}{|c|c|cccc|cccc|cccc|cccc|}
        \hline
        & & \multicolumn{4}{c|}{GPT-4o-mini} & \multicolumn{4}{c|}{GPT-4o}  & \multicolumn{4}{c|}{Llama3.1-8b} & \multicolumn{4}{c|}{Llama3.1-70b} \\
        \hline
        & Method & \precision & \recall & \fscore & \consistency & \precision & \recall & \fscore & \consistency& \precision & \recall & \fscore & \consistency & \precision & \recall & \fscore & \consistency \\
        \hhline{|==================|}
        \multirow{3}{*}{PAGED} & \greedy & 77.88 & 74.17 & 75.19 & 95.40 & 81.01 & 78.68 & 79.13 & 96.63 & 76.87 & 78.42 & 76.98 & 94.48 & 79.28 & 81.15 & 79.54 & 93.25 \\
        & \mv & \textbf{80.10} & 70.32 & 73.93 & 66.26 & \textbf{82.90} & 75.62 & 78.17 & 71.47 & \textbf{79.87} & 70.02 & 73.43 & 51.23 & \textbf{80.01} & 78.67 & 78.66 & 69.33 \\
        & \approach & 77.81 & \textbf{76.88} & \textbf{76.59} & \textbf{99.08} & 80.87 & \textbf{79.93} & \textbf{79.73} & \textbf{99.08} & 77.47 & \textbf{79.31} & \textbf{77.79} & \textbf{98.47} & 79.13 & \textbf{81.69} & \textbf{79.85} & \textbf{96.63}  \\
        \hhline{|==================|}
        \multirow{3}{*}{WordNet} & \greedy & 72.53 & 52.80 & 59.20 &  78.00 & 75.84 & 66.58 & 69.69 & 83.00 & 67.24 & 55.55 & 59.84 &  65.00 & 78.56 & 66.97 & 71.14 &  95.00 \\
        & \mv & \textbf{83.06} & 43.42 & 54.28 &  64.00 & \textbf{84.13} & 54.48 & 63.97 & 75.00 & \textbf{82.23} & 33.16 & 44.55 &  65.00 & \textbf{86.24} & 52.65 & 63.13 &  80.00 \\
        & \approach & 82.91 & \textbf{69.22} & \textbf{73.83} & \textbf{100} & 80.01 & \textbf{73.52} & \textbf{75.93} & \textbf{99.00} & 74.99 & \textbf{64.75} & \textbf{68.81} & \textbf{100} & 80.24 & \textbf{73.42} & \textbf{75.94} & \textbf{100} \\
        \hhline{==================}
         & Method & \multicolumn{2}{c}{\accuracy} & \multicolumn{2}{c|}{\sr} & \multicolumn{2}{c}{\accuracy} & \multicolumn{2}{c|}{\sr} & \multicolumn{2}{c}{\accuracy} & \multicolumn{2}{c|}{\sr} & \multicolumn{2}{c}{\accuracy} & \multicolumn{2}{c|}{\sr} \\
         \hhline{|==================|}
        \multirow{5}{*}{Clevr} & \greedy & \multicolumn{2}{c}{39.00} & \multicolumn{2}{c|}{45.67} & \multicolumn{2}{c}{65.33} &  \multicolumn{2}{c|}{71.33} & \multicolumn{2}{c}{38.00} &  \multicolumn{2}{c|}{48.33} & \multicolumn{2}{c}{65.00} &  \multicolumn{2}{c|}{72.00} \\
        & \mv & \multicolumn{2}{c}{21.00} & \multicolumn{2}{c|}{65.67} & \multicolumn{2}{c}{51.67} &  \multicolumn{2}{c|}{77.33} & \multicolumn{2}{c}{19.00} &  \multicolumn{2}{c|}{86.33} & \multicolumn{2}{c}{57.67} &  \multicolumn{2}{c|}{90.33} \\
        & \esc & \multicolumn{2}{c}{33.67} & \multicolumn{2}{c|}{39.00} & \multicolumn{2}{c}{66.00} &  \multicolumn{2}{c|}{71.00} & \multicolumn{2}{c}{28.67} &  \multicolumn{2}{c|}{31.67} & \multicolumn{2}{c}{70.00} &  \multicolumn{2}{c|}{76.00} \\
        & \escf & \multicolumn{2}{c}{65.33} & \multicolumn{2}{c|}{80.33} & \multicolumn{2}{c}{80.33} & \multicolumn{2}{c|}{86.67} & \multicolumn{2}{c}{73.00} & \multicolumn{2}{c|}{94.67} & \multicolumn{2}{c}{88.33} & \multicolumn{2}{c|}{96.00} \\
        & \approach & \multicolumn{2}{c}{\textbf{69.67}} & \multicolumn{2}{c|}{\textbf{98.33}} & \multicolumn{2}{c}{\textbf{81.33}} & \multicolumn{2}{c|}{\textbf{100}} & \multicolumn{2}{c}{\textbf{74.67}} & \multicolumn{2}{c|}{\textbf{100}} & \multicolumn{2}{c}{\textbf{89.67}} & \multicolumn{2}{c|}{\textbf{100}} \\
        \hline
    \end{tabular}
    }
\end{table*}

\paragraph{Rationale and setup}
In this research question, we evaluate the effectiveness of \approach{} from two perspectives: quality and consistency, comparing it against baseline approaches across various types of graph models. Additionally, we assess performance across \llms{} of different sizes to understand the impact of \llms{} on the approach. For each dataset sample, 10 candidates are generated for abstraction.

\paragraph{Effectiveness results}
Evaluation results for different approaches using various \llms{} across multiple datasets are presented in \autoref{tab:rq1}. In terms of model consistency, \approach{} significantly improves the consistency of generated models with respect to well-formedness constraints. Notably, \approach{} produces consistent models for \emph{all samples} in 6 out of 12 cases. In the remaining cases, \approach{} remains consistently above $96.6\%$. We suspect that the small fraction of inconsistent models may result from the inherent limitations of LLMs in generating consistent models for certain descriptions.

The improvement in consistency helps with enhanced model quality. Overall, \approach{} outperforms \greedy{} generation across all datasets. The F1-score improves by an average of $0.78\%$ for the PAGED dataset and $8.61\%$ for WordNet. For the Clevr dataset, since a model needs to be consistent to derive a valid answer, answer accuracy increases by approximately $27\%$ on average with \approach{}. 

Compared to the baselines, for non-executable models, recall improves by $0.55\% - 2.69\%$ on the PAGED dataset and $6.45\% - 16.42\%$ on WordNet. Compared to the \mv{} baseline, \approach{} generally achieves slightly lower precision. However, the higher precision of \mv{} comes at the cost of a significant reduction in recall, leading to \emph{overall lower} F1-scores than the \greedy{} approach. In contrast, \approach{} improves recall while preserving precision, resulting in a higher F1-score compared to both the \greedy{} and \mv{} approaches.

For executable models, both \mv{} and \esc{} suffer from low success rates, leading to worse answer accuracy than \greedy{}. Filtering out inconsistent models allows \escf{} to significantly improve accuracy. However, execution-based self-consistency treats each candidate model independently. By capturing correct subgraphs across candidates, \approach{} further improves accuracy by $1.00\% - 4.33\%$ compared to \escf{}.


\begin{tcolorbox}[boxsep=-1mm]
    \textbf{RQ1.1.}
    Models generated by \approach{} exhibit significantly higher consistency than all baselines, achieving perfect consistency in 6 out of 12 cases. This improvement also leads to better model quality compared to baseline approaches, particularly \greedy{} generation, with average F1-score gains ranging from $0.78\%$ to $27\%$. Notably, \approach{} even outperforms the \escf{} baseline, which is specifically designed for executable models.
\end{tcolorbox}

\paragraph{Impact of LLM sizes}
Comparing the performance of \greedy{} generation across different LLM sizes in \autoref{tab:rq1}, larger models (GPT-4o and Llama3.1-70B) consistently outperform smaller ones (GPT-4o-mini and Llama3.1-8B). Interestingly, while \approach{} improves the generation performance for all \llms{} compared to \greedy{}, the improvements achieved are more pronounced for smaller LLMs than for larger ones. For example, on the WordNet dataset, \approach{} improves the F1-score of GPT-4o-mini by $14.64\%$, whereas the improvement for GPT-4o is only $6.24\%$. Similarly, accuracy on the Clevr dataset increases by over $36\%$ for Llama3.1-8B, compared to approximately $25\%$ for Llama3.1-70B.

Due to this greater impact on models generated by smaller \llms{}, \approach{} can produce higher-quality models with smaller LLMs compared with \greedy{} generation with larger LLMs. In half of the cases, models generated by \approach{} using GPT-4o-mini and Llama3.1-8B outperform those produced by their larger counterparts via \greedy{} generation. This improvement highlights the potential of \approach{} in resource-constrained environments where only small LLMs can be used.

\begin{tcolorbox}[boxsep=-1mm]
    \textbf{RQ1.2.}
    While larger LLMs produce higher-quality models than smaller LLMs, the improvement achieved by \approach{} is more pronounced for smaller \llms{}. In 3 out of 6 cases, smaller \llms{} combined with \approach{} outperform their larger counterparts using \greedy{}.
\end{tcolorbox}

\subsection{RQ2: Impact of consistency on model quality}

\paragraph{Rationale and setup}
In RQ1, we demonstrate that \approach{} generates more consistent and accurate models than baseline approaches. Notably, when compared to \mv{}, the improvement in consistency leads to a significant increase in model quality. In this RQ, we examine how consistency affects the quality of a generated model and explore the extent to which a consistency guarantee can enhance model quality.

To examine this influence, we use the same generated candidate models from RQ1, categorizing them into two groups: consistent and inconsistent models. The average model quality score in each group is computed over 10 runs. Since the performance distribution of LLM-generated models is unknown, 
we use the non-parametric Wilcoxon rank-sum test to assess statistical significance. Additionally, we use Cliff's Delta, a non-parametric effect size measure, to quantify the impact of consistency on model quality.



\paragraph{Results}
\begin{figure*}
    \centering
    \includegraphics[width=0.28\textwidth]{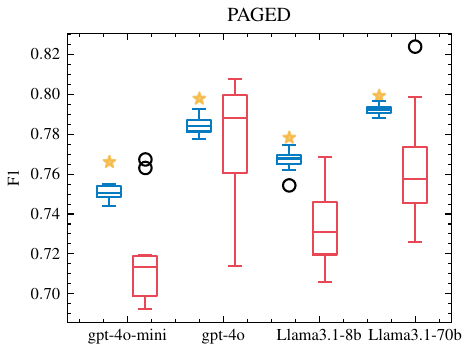}
    \includegraphics[width=0.28\textwidth]{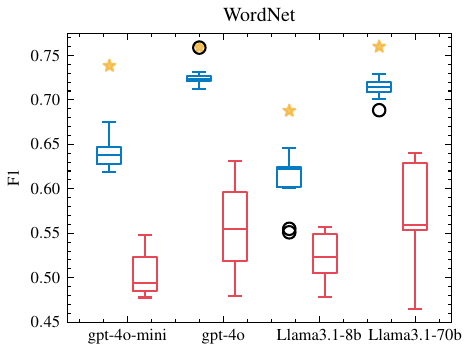}
    \includegraphics[width=0.28\textwidth]{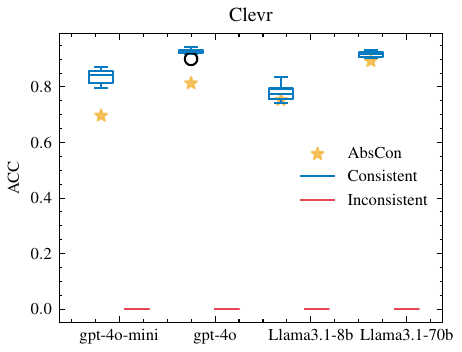}
    \caption{Distribution of average quality of consistent and inconsistent models generated by LLMs}
    \label{fig:consistency}
\end{figure*}

\autoref{fig:consistency} presents a box plot of the average quality metric values for consistent and inconsistent models generated by \llms{}. Overall, consistent models exhibit notably higher quality than inconsistent ones, with the most notable difference observed in the Clevr dataset. Since these models must be executed to produce a final answer, an inconsistent model always results in an execution error, leading to zero accuracy. Consistent models also demonstrate better F1-score in the PAGED and WordNet datasets, except for GPT-4o on the PAGED dataset, where this trend does not hold.

Statistical tests are then conducted to determine whether such differences are significant, with the alternative hypothesis stating that the average quality of consistent models is higher. The statistical test for the Clevr dataset is excluded since the scores of inconsistent models are always zero. The results reject the null hypothesis in all cases except for GPT-4o on the PAGED dataset, with $p \leq 0.02$. For WordNet, the null hypothesis is rejected across all cases, with $p \leq 0.001$. We further assess the effect size, finding that, except for GPT-4o on the PAGED dataset, all effect sizes exceed $0.6$, indicating a large impact of consistency on model quality \cite{meissel2024using}.

We also plot the average quality of \textit{all} models generated by \approach{}. Since most models produced by \approach{} are consistent, their quality is generally comparable to that of consistent models in the candidates. This observation suggests that the consistency guarantee provided by \approach{} effectively enhances model quality. In the PAGED and WordNet datasets, models from \approach{} outperform even the consistent model candidates. We attribute this improvement to the advantage of considering multiple candidates in \approach{}, which helps correct errors made by \llms{} when producing a single candidate.

\begin{tcolorbox}[boxsep=-1mm]
    \textbf{RQ2.}
    Consistent models generated by LLMs exhibit significantly higher quality than inconsistent models. The consistency guarantee provided by \approach{} effectively enhances model quality, with models from \approach{} surpassing the quality of consistent models from the candidates in 8 out of 12 cases.
\end{tcolorbox}

\subsection{RQ3: Impact of number of candidates}
\begin{figure*}
    \centering
    \includegraphics[width=0.3\textwidth]{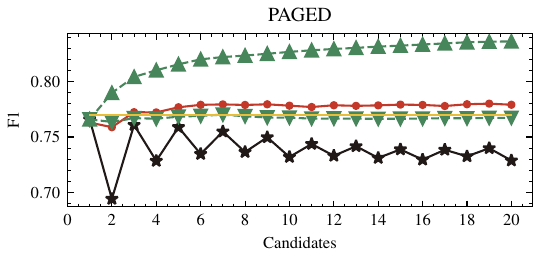}
    \includegraphics[width=0.3\textwidth]{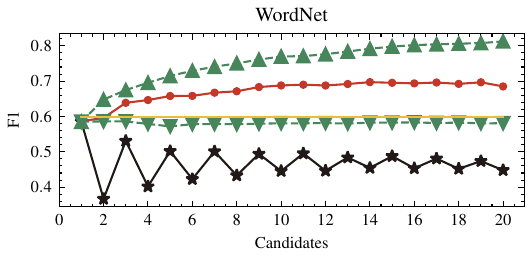}
    \includegraphics[width=0.3\textwidth]{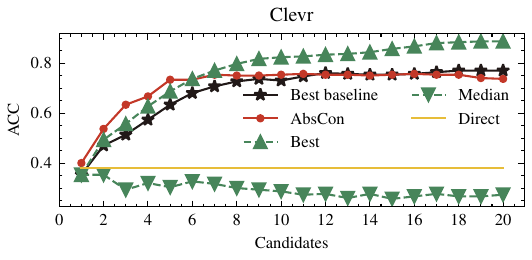}
    \includegraphics[width=0.3\textwidth]{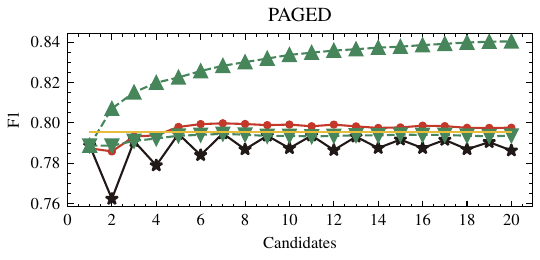}
    \includegraphics[width=0.3\textwidth]{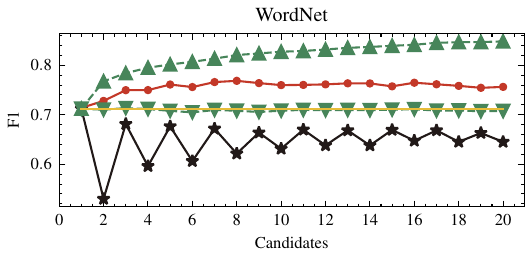}
    \includegraphics[width=0.3\textwidth]{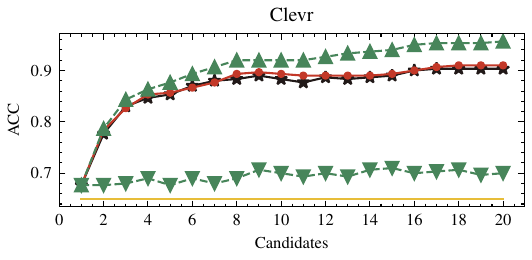}
    \caption{Impact of number of candidates on the proposed method \approach{} on Llama3.1-8b (top) and Llama3.1-70b (bottom); impacts of candidates on the best baseline (\mv{} for PAGED and WordNet, \escf{} for Clevr), \greedy{}, and two oracle evaluators (median and best) are also shown}
    \label{fig:candidates}
\end{figure*}

\paragraph{Rationale and setup}
One crucial parameter in \approach{} is the number of candidates. The purpose of constructing the partial model in \approach{} is to estimate the distribution of models an LLM can generate from a given input. Naturally, one might hypothesize that increasing the number of candidates would improve this estimation, thereby enhancing the quality of the concrete models. In this research question, we empirically evaluate this hypothesis.

In this experiment, we analyze how quality metrics change with the candidate counts, ranging from 1 to 20.  We present results for two LLMs from the Llama3.1 family to assess the impact of candidate count under the same LLM architecture but with different model sizes. Quality scores are plotted for \greedy{}, \approach{}, and the best-performing baseline (\mv{} for PAGED and WordNet, \escf{} for Clevr).

\paragraph{Results}
The trend of quality scores with respect to the number of candidates is shown in \autoref{fig:candidates}. In general, quality scores increase as the number of candidates grows, though the magnitude of improvement varies across datasets. A larger candidate set increases the likelihood of including correct elements, thereby enhancing the overall quality. This improvement plateaus as the number of candidates increases, stabilizing at around 5 candidates for Llama3.1-8B and between 5 and 8 candidates for Llama3.1-70B. This result suggests that only a small number of candidates is needed to achieve significant improvements over the \greedy{} approach. For Llama3.1-8B on WordNet and Clevr, performance slightly declines as the number of candidates increases. We suspect this is due to smaller LLMs tending to repeat common mistakes, making these errors more dominant among the candidates.

Compared to other approaches, \approach{} consistently produces higher-quality models than both the \greedy{} and baseline methods. To assess how well \approach{} performs relative to the best possible outcome, assuming an oracle evaluator is available, we also plot the median and best metric scores of the candidates. Note that these best candidates are difficult to identify in practice due to lack of oracle evaluator. Generally, models generated by \approach{} outperform the median candidate performance. Notably, for Llama3.1-8B on the Clevr dataset, \approach{} even surpasses the best candidate when using five candidates. Moreover, while individual candidates lack a consistency guarantee, \approach{} ensures consistency through concretization. However, the gap between \approach{} and the best candidate suggests room for further improvement, potentially by refining the abstraction and concretization processes.

\begin{tcolorbox}[boxsep=-1mm]
    \textbf{RQ3.}
    Model quality improves as the number of candidates increases. The gains plateau quickly at around 5 to 8 candidates, suggesting that \approach{} requires only a few candidates to be effective. However, there is still room for improvement compared to the best possible model.
\end{tcolorbox}

\subsection{Discussion}
\noindent\textbf{Constraints.}
In \emph{RQ1}, \approach{} and \mv{} combine graph models using self-consistency, with and without constraints. While \mv{} achieves slightly higher precision, \approach{} consistently yields much better recall and F1-scores for all LLMs and datasets. This finding highlights that \textbf{constraints are crucial for graph model generation}, particularly in preserving recall. Since ground truth models are typically consistent, constraints serve as heuristics for capturing correct elements in the partial model during concretization to create a better output.



Furthermore, \emph{RQ2} reveals that the quality of consistent models is significantly higher than that of inconsistent ones, emphasizing the importance of constraints in generation. In practice, such constraints can often be specified by domain experts or automatically derived from existing models \cite{rabbani2023extraction}.

\noindent\textbf{Candidates.}
\emph{RQ3} shows that increasing the number of candidates generally improves the quality of the final output model. While generating more candidates increases both time and computational cost, each candidate is independent and can be generated in parallel. Moreover, only a few candidates are needed to achieve significant improvements. Candidate quality is also crucial; common mistakes among candidates can propagate, leading to incorrect final models. Additionally, candidate diversity may also have an impact on the performance and remains a promising direction for future investigation.



\noindent\textbf{Applicability.}
While \approach{} is designed to address uncertainty in \llms{} and enhance model quality, it can also be applied to other scenarios by varying the candidate generation process. For example, candidates could be generated by multiple domain experts or by combining \llm{} outputs with human-created solutions. However, a key assumption of this approach is that candidates share common, correct subgraphs.

Additionally, different abstraction and concretization methods can be adapted based on the domain. For instance, the objective function for concretization could be optimized using alternative criteria, such as minimum description length \cite{nair2024midgard}.

\approach{} acts as an inference-time compute improvement~\cite{snell2024scaling}, using extra computation (abstraction and concretization) to boost LLM performance without retraining. Most test samples are processed within seconds. We leave the detailed study on the runtime cost of \approach{} to the future. 


\subsection{Threats to validity}
\noindent\textbf{Internal validity.}
The output of LLMs can vary for the same input, depending on the temperature setting. 
This work leverages such variations by merging multiple candidates generated by LLMs. LLM performance may vary based on the prompting technique. To mitigate these variations, we follow best practices such as chain-of-thought and few-shot prompting. We also conduct evaluations across multiple LLMs and tasks, obtaining consistent results.

\noindent\textbf{External validity.}
The effect of \approach{} may vary for different domains and model types. To systematically investigate the effectiveness of \approach{}. We use a widely adopted taxonomy dataset for structural models. However, due to the limited availability of datasets for behavioral and executable models, we adapt similar datasets from other domains, transforming them into behavioral and executable models.

\noindent\textbf{Construct validity.}
Evaluating model quality is inherently challenging. To ensure accurate evaluations, we select appropriate metrics based on the model type. We use standard precision, recall, and F1-scores when node labels are fixed, soft metrics when labels may vary, and execution accuracy for executable models, aligning with previous studies \cite{chen2022consistent,chen2023prompting}.


\section{Related Work}

\noindent\textbf{Uncertainty and consistent model generation.}
Modeling uncertainty in graphs is an active research area. Partial models \cite{famelis2012partial} and 150\% models explicitly represent uncertainty through annotations in graphs. 
These approaches have been widely applied, including model checking \cite{bauer2013weighted}, requirements engineering \cite{salay2013managing}, and consistent model generation \cite{semerath2021automated,marussy2020specification}.

Logic-based solvers
are widely used for consistent graph model generation \cite{soltana2020practical}, but they often face scalability challenges. To improve scalability, hybrid methods combining solvers with meta-heuristic search have been proposed \cite{semerath2021automated,semerath2018graph}. Refinery \cite{marussy2024refinery} provides a web-based tool for generating consistent models from a metamodel and constraints.

In this paper, we enhance consistent graph model generation using \llms{} by explicitly modeling the uncertainty inherent in \llms{} through a probabilistic partial model. Unlike conventional model generation approaches, our method generates models directly from natural language descriptions.




\noindent\textbf{Self-consistency for LLMs.}
Self-consistency was initially proposed as a strategy that samples multiple reasoning paths and selects the most frequent answer to improve the performance of \llms{} \cite{wang2023self}. 
However, the original method is limited to tasks with a fixed set of possible answers. Universal self-consistency \cite{chen2023universal} extends this approach by leveraging LLMs to determine the most frequent response, while atomic self-consistency \cite{thirukovalluru2024atomic} further refines the method by decomposing each answer into multiple atomic elements. Additionally, reasoning-aware self-consistency \cite{wan2024dynamic} assesses consistency across the reasoning process and the final answer.

MIDGARD \cite{nair2024midgard} presents an alternative self-consistency technique for graphs. However, MIDGARD does not explicitly capture matching elements among candidates and is restricted to directed acyclic graphs. In contrast, \approach{} explicitly models uncertainty, enabling different concretization techniques. Moreover, our proposed concretization method generates consistent models for any well-formedness constraints expressible in the chosen constraint language.

\noindent\textbf{LLMs and MDE.}
LLMs have gained increasing interest in MDE, as highlighted by Di Rocco et al. in a survey \cite{di2025use}.

In modeling tasks, significant attention has been given to generating models from textual descriptions. LLMs have been applied to generate various types of models, including domain models \cite{yang2024multi,camara2023assessment,chen2023automated}, goal models \cite{chen2023use}, and sequence diagrams \cite{ferrari2024model,jahan2024automated}. Additionally, they have been used to detect inconsistencies across models \cite{sultan2024ai} and to translate natural language into modeling query languages \cite{lopez2024text2vql}.

This work falls within the category of using LLMs for model generation from natural language descriptions. Moreover, it leverages modeling techniques to explicitly capture the uncertainty of LLM-generated candidates, ensuring accurate and consistent models. While this paper focuses on specific model generation tasks, \approach{} is also applicable to other types of models, such as domain and goal models. These alternatives are excluded due to the lack of large datasets and challenges in automated evaluation \cite{chen2023automated}. Extending \approach{} to these areas remains a promising direction for future work.
\section{Conclusion and future work}
This paper proposes \approach{}, a framework that applies self-consistency to \llm{}-generated models, enhancing model quality while ensuring compliance with constraints. We instantiate this framework using graph-similarity-based abstraction and constraint optimization-based concretization. \approach{} contributes to two key challenges in model generation from textual description using LLMs: (1) it guarantees the \textit{consistency} of the model, and (2) it effectively combines partially correct solutions to produce a more \textit{accurate} model. We evaluate \approach{} on both open-source and closed-source LLMs, including Llama3.1 and GPT-4o, across diverse model generation tasks.
The results demonstrate that \approach{} significantly improves the consistency of generated models and enhances overall quality across various metrics. However, there is still room for improvement compared to the best possible model. 

In future work, we aim to explore the performance of \approach{} on reasoning \llms{} such as DeepSeek R1 \cite{guo2025deepseek}. We also plan to investigate the impact of alternative objective functions and candidate diversity, extend our approach to support high-level constraint languages, and apply it to more complex models such as domain models.

\bibliographystyle{IEEEtran}
\bibliography{sample-base}

\begin{thebibliography}{10}
\providecommand{\url}[1]{#1}
\csname url@samestyle\endcsname
\providecommand{\newblock}{\relax}
\providecommand{\bibinfo}[2]{#2}
\providecommand{\BIBentrySTDinterwordspacing}{\spaceskip=0pt\relax}
\providecommand{\BIBentryALTinterwordstretchfactor}{4}
\providecommand{\BIBentryALTinterwordspacing}{\spaceskip=\fontdimen2\font plus
\BIBentryALTinterwordstretchfactor\fontdimen3\font minus \fontdimen4\font\relax}
\providecommand{\BIBforeignlanguage}[2]{{%
\expandafter\ifx\csname l@#1\endcsname\relax
\typeout{** WARNING: IEEEtran.bst: No hyphenation pattern has been}%
\typeout{** loaded for the language `#1'. Using the pattern for}%
\typeout{** the default language instead.}%
\else
\language=\csname l@#1\endcsname
\fi
#2}}
\providecommand{\BIBdecl}{\relax}
\BIBdecl

\bibitem{achiam2023gpt}
J.~Achiam, S.~Adler, S.~Agarwal, L.~Ahmad, I.~Akkaya, F.~L. Aleman, D.~Almeida, J.~Altenschmidt, S.~Altman, S.~Anadkat \emph{et~al.}, ``{GPT-4} technical report,'' \emph{arXiv preprint arXiv:2303.08774}, 2023.

\bibitem{dubey2024llama}
A.~Dubey, A.~Jauhri, A.~Pandey, A.~Kadian, A.~Al-Dahle, A.~Letman, A.~Mathur, A.~Schelten, A.~Yang, A.~Fan \emph{et~al.}, ``The llama 3 herd of models,'' \emph{arXiv preprint arXiv:2407.21783}, 2024.

\bibitem{guo2025deepseek}
D.~Guo, D.~Yang, H.~Zhang, J.~Song, R.~Zhang, R.~Xu, Q.~Zhu, S.~Ma, P.~Wang, X.~Bi \emph{et~al.}, ``{DeepSeek-R1}: Incentivizing reasoning capability in {LLMs} via reinforcement learning,'' \emph{arXiv preprint arXiv:2501.12948}, 2025.

\bibitem{di2025use}
J.~Di~Rocco, D.~Di~Ruscio, C.~Di~Sipio, P.~T. Nguyen, and R.~Rubei, ``On the use of large language models in model-driven engineering,'' \emph{Software and Systems Modeling}, pp. 1--26, 2025.

\bibitem{camara2023assessment}
J.~C{\'a}mara, J.~Troya, L.~Burgue{\~n}o, and A.~Vallecillo, ``{On the assessment of generative AI in modeling tasks: an experience report with ChatGPT and UML},'' \emph{Software and Systems Modeling}, vol.~22, no.~3, pp. 781--793, 2023.

\bibitem{chen2023automated}
K.~Chen, Y.~Yang, B.~Chen, J.~A.~H. L{\'o}pez, G.~Mussbacher, and D.~Varr{\'o}, ``Automated domain modeling with large language models: A comparative study,'' in \emph{2023 ACM/IEEE 26th International Conference on Model Driven Engineering Languages and Systems (MODELS)}.\hskip 1em plus 0.5em minus 0.4em\relax IEEE, 2023, pp. 162--172.

\bibitem{arulmohan2023extracting}
S.~Arulmohan, M.-J. Meurs, and S.~Mosser, ``Extracting domain models from textual requirements in the era of large language models,'' in \emph{2023 ACM/IEEE International Conference on Model Driven Engineering Languages and Systems Companion (MODELS-C)}.\hskip 1em plus 0.5em minus 0.4em\relax IEEE, 2023, pp. 580--587.

\bibitem{chen2023use}
B.~Chen, K.~Chen, S.~Hassani, Y.~Yang, D.~Amyot, L.~Lessard, G.~Mussbacher, M.~Sabetzadeh, and D.~Varr{\'o}, ``On the use of {GPT-4} for creating goal models: An exploratory study,'' in \emph{2023 IEEE 31st International Requirements Engineering Conference Workshops (REW)}.\hskip 1em plus 0.5em minus 0.4em\relax IEEE, 2023, pp. 262--271.

\bibitem{ferrari2024model}
A.~Ferrari, S.~Abualhaija, and C.~Arora, ``Model generation with {LLM}s: From requirements to {UML} sequence diagrams,'' in \emph{2024 IEEE 32nd International Requirements Engineering Conference Workshops (REW)}, 2024, pp. 291--300.

\bibitem{chen2023prompting}
B.~Chen, F.~Yi, and D.~Varr{\'o}, ``Prompting or fine-tuning? a comparative study of large language models for taxonomy construction,'' in \emph{2023 ACM/IEEE International Conference on Model Driven Engineering Languages and Systems Companion (MODELS-C)}.\hskip 1em plus 0.5em minus 0.4em\relax IEEE, 2023, pp. 588--596.

\bibitem{varro2018towards}
D.~Varr{\'o}, O.~Semer{\'a}th, G.~Sz{\'a}rnyas, and {\'A}.~Horv{\'a}th, ``Towards the automated generation of consistent, diverse, scalable and realistic graph models,'' in \emph{Graph Transformation, Specifications, and Nets: In Memory of Hartmut Ehrig}.\hskip 1em plus 0.5em minus 0.4em\relax Springer, 2018, pp. 285--312.

\bibitem{garzon2015umple}
M.~A. Garz{\'o}n, H.~Aljamaan, and T.~C. Lethbridge, ``Umple: A framework for model driven development of object-oriented systems,'' in \emph{2015 IEEE 22nd International Conference on Software Analysis, Evolution, and Reengineering (SANER)}.\hskip 1em plus 0.5em minus 0.4em\relax IEEE, 2015, pp. 494--498.

\bibitem{gurbuz2018model}
H.~G. Gurbuz and B.~Tekinerdogan, ``Model-based testing for software safety: a systematic mapping study,'' \emph{Software Quality Journal}, vol.~26, no.~4, pp. 1327--1372, 2018.

\bibitem{plantuml}
``Plantuml at a glance,'' \url{https://plantuml.com/}.

\bibitem{netz2024using}
L.~Netz, J.~Reimar, and B.~Rumpe, ``Using grammar masking to ensure syntactic validity in {LLM}-based modeling tasks,'' in \emph{Proceedings of the {ACM/IEEE} 27th International Conference on Model Driven Engineering Languages and Systems Companion, ({MODELS-C})}.\hskip 1em plus 0.5em minus 0.4em\relax {ACM}, 2024, pp. 115--122.

\bibitem{mousavi2024towards}
\BIBentryALTinterwordspacing
J.~Mousavi and A.~Termehchy, ``Towards consistent language models using controlled prompting and decoding,'' in \emph{Neuro-Symbolic Learning and Reasoning in the era of Large Language Models}, 2024. [Online]. Available: \url{https://openreview.net/forum?id=UQTGlkr30Y}
\BIBentrySTDinterwordspacing

\bibitem{wang2023self}
\BIBentryALTinterwordspacing
X.~Wang, J.~Wei, D.~Schuurmans, Q.~V. Le, E.~H. Chi, S.~Narang, A.~Chowdhery, and D.~Zhou, ``Self-consistency improves chain of thought reasoning in language models,'' in \emph{The Eleventh International Conference on Learning Representations}.\hskip 1em plus 0.5em minus 0.4em\relax OpenReview.net, 2023. [Online]. Available: \url{https://openreview.net/forum?id=1PL1NIMMrw}
\BIBentrySTDinterwordspacing

\bibitem{wei2022chain}
J.~Wei, X.~Wang, D.~Schuurmans, M.~Bosma, B.~Ichter, F.~Xia, E.~H. Chi, Q.~V. Le, and D.~Zhou, ``Chain-of-thought prompting elicits reasoning in large language models,'' \emph{Advances in Neural Information Processing Systems}, vol.~35, pp. 24\,824--24\,837, 2022.

\bibitem{openai2024learning}
\BIBentryALTinterwordspacing
OpenAI. (2024) Learning to reason with {LLMs}. Accessed on 2024-10-31. [Online]. Available: \url{https://openai.com/index/learning-to-reason-with-llms}
\BIBentrySTDinterwordspacing

\bibitem{snell2024scaling}
\BIBentryALTinterwordspacing
C.~V. Snell, J.~Lee, K.~Xu, and A.~Kumar, ``Scaling {LLM} test-time compute optimally can be more effective than scaling parameters for reasoning,'' in \emph{The Thirteenth International Conference on Learning Representations}, 2025. [Online]. Available: \url{https://openreview.net/forum?id=4FWAwZtd2n}
\BIBentrySTDinterwordspacing

\bibitem{saini2022machine}
R.~Saini, G.~Mussbacher, J.~L. Guo, and J.~Kienzle, ``Machine learning-based incremental learning in interactive domain modelling,'' in \emph{Proceedings of the 25th International Conference on Model Driven Engineering Languages and Systems}, 2022, pp. 176--186.

\bibitem{semerath2021automated}
O.~Semer{\'a}th, A.~A. Babikian, B.~Chen, C.~Li, K.~Marussy, G.~Sz{\'a}rnyas, and D.~Varr{\'o}, ``Automated generation of consistent, diverse and structurally realistic graph models,'' \emph{Software and Systems Modeling}, vol.~20, no.~5, pp. 1713--1734, 2021.

\bibitem{zhu2022survey}
Y.~Zhu, Y.~Du, Y.~Wang, Y.~Xu, J.~Zhang, Q.~Liu, and S.~Wu, ``A survey on deep graph generation: Methods and applications,'' in \emph{Learning on Graphs Conference}.\hskip 1em plus 0.5em minus 0.4em\relax PMLR, 2022, pp. 47--1.

\bibitem{famelis2012partial}
M.~Famelis, R.~Salay, and M.~Chechik, ``Partial models: Towards modeling and reasoning with uncertainty,'' in \emph{2012 34th International Conference on Software Engineering (ICSE)}.\hskip 1em plus 0.5em minus 0.4em\relax IEEE, 2012, pp. 573--583.

\bibitem{salay2012language}
R.~Salay, M.~Famelis, and M.~Chechik, ``Language independent refinement using partial modeling,'' in \emph{International Conference on Fundamental Approaches to Software Engineering}.\hskip 1em plus 0.5em minus 0.4em\relax Springer, 2012, pp. 224--239.

\bibitem{famelis2011partial}
M.~Famelis, S.~Ben-David, M.~Chechik, and R.~Salay, ``Partial models: A position paper,'' in \emph{Proceedings of the 8th International Workshop on Model-Driven Engineering, Verification and Validation}, 2011, pp. 1--4.

\bibitem{marussy2020specification}
K.~Marussy, O.~Semer{\'a}th, A.~A. Babikian, and D.~Varr{\'o}, ``A specification language for consistent model generation based on partial models,'' \emph{The Journal of Object Technology}, vol.~19, no.~3, pp. 3--1, 2020.

\bibitem{semerath2018graph}
O.~Semer{\'a}th, A.~S. Nagy, and D.~Varr{\'o}, ``A graph solver for the automated generation of consistent domain-specific models,'' in \emph{Proceedings of the 40th International Conference on Software Engineering}, 2018, pp. 969--980.

\bibitem{chen2023universal}
\BIBentryALTinterwordspacing
X.~Chen, R.~Aksitov, U.~Alon, J.~Ren, K.~Xiao, P.~Yin, S.~Prakash, C.~Sutton, X.~Wang, and D.~Zhou, ``Universal self-consistency for large language models,'' in \emph{ICML 2024 Workshop on In-Context Learning}, 2024. [Online]. Available: \url{https://openreview.net/forum?id=LjsjHF7nAN}
\BIBentrySTDinterwordspacing

\bibitem{thirukovalluru2024atomic}
R.~Thirukovalluru, Y.~Huang, and B.~Dhingra, ``Atomic self-consistency for better long form generations,'' in \emph{Proceedings of the 2024 Conference on Empirical Methods in Natural Language Processing}.\hskip 1em plus 0.5em minus 0.4em\relax Association for Computational Linguistics, 2024, pp. 12\,681--12\,694.

\bibitem{forrest2024cbc}
\BIBentryALTinterwordspacing
{J. Forrest, et. al.}, ``coin-or/cbc: Release releases/2.10.12,'' 2024. [Online]. Available: \url{https://doi.org/10.5281/zenodo.13347261}
\BIBentrySTDinterwordspacing

\bibitem{holmstrom2009user}
K.~Holmstr{\"o}m, A.~O. G{\"o}ran, and M.~M. Edvall, ``User’s guide for tomlab/cplex v12. 1,'' \emph{Tomlab Optim. Retrieved}, vol.~1, p. 2017, 2009.

\bibitem{gurobi2024}
\BIBentryALTinterwordspacing
{Gurobi Optimization, LLC}, ``Gurobi optimizer reference manual,'' 2024. [Online]. Available: \url{https://www.gurobi.com}
\BIBentrySTDinterwordspacing

\bibitem{knut2024mermaid}
\BIBentryALTinterwordspacing
K.~Sveidqvist and {Contributors to Mermaid}, ``{Mermaid: Generate diagrams from markdown-like text},'' Dec. 2014. [Online]. Available: \url{https://github.com/mermaid-js/mermaid}
\BIBentrySTDinterwordspacing

\bibitem{soltana2020practical}
G.~Soltana, M.~Sabetzadeh, and L.~C. Briand, ``Practical constraint solving for generating system test data,'' \emph{ACM Transactions on Software Engineering and Methodology (TOSEM)}, vol.~29, no.~2, pp. 1--48, 2020.

\bibitem{marussy2024refinery}
K.~Marussy, A.~Ficsor, O.~Semer{\'a}th, and D.~Varr{\'o}, ``Refinery: Graph solver as a service: Refinement-based generation and analysis of consistent models,'' in \emph{Proceedings of the 2024 IEEE/ACM 46th International Conference on Software Engineering: Companion Proceedings}, 2024, pp. 64--68.

\bibitem{chen2024embedding}
K.~Chen, B.~Chen, Y.~Yang, G.~Mussbacher, and D.~Varr{\'o}, ``Embedding-based automated assessment of domain models,'' in \emph{Proceedings of the ACM/IEEE 27th International Conference on Model Driven Engineering Languages and Systems Companion ({MODELS-C})}, 2024, pp. 87--94.

\bibitem{abu2015exact}
Z.~Abu-Aisheh, R.~Raveaux, J.-Y. Ramel, and P.~Martineau, ``An exact graph edit distance algorithm for solving pattern recognition problems,'' in \emph{4th International Conference on Pattern Recognition Applications and Methods 2015}, 2015, pp. 271--278.

\bibitem{bergmann2014translating}
G.~Bergmann, ``Translating {OCL} to graph patterns,'' in \emph{International Conference on Model Driven Engineering Languages and Systems}.\hskip 1em plus 0.5em minus 0.4em\relax Springer, 2014, pp. 670--686.

\bibitem{semerath2017formal}
O.~Semer{\'a}th, {\'A}.~Barta, {\'A}.~Horv{\'a}th, Z.~Szatm{\'a}ri, and D.~Varr{\'o}, ``Formal validation of domain-specific languages with derived features and well-formedness constraints,'' \emph{Software and Systems Modeling}, vol.~16, pp. 357--392, 2017.

\bibitem{chen2022consistent}
B.~Chen, K.~Marussy, S.~Pilarski, O.~Semer{\'a}th, and D.~Varr{\'o}, ``Consistent scene graph generation by constraint optimization,'' in \emph{Proceedings of the 37th IEEE/ACM International Conference on Automated Software Engineering}, 2022, pp. 1--13.

\bibitem{cocos2018comparing}
A.~Cocos, M.~Apidianaki, and C.~Callison-Burch, ``Comparing constraints for taxonomic organization,'' in \emph{Proceedings of the 2018 Conference of the North American Chapter of the Association for Computational Linguistics: Human Language Technologies, Volume 1 (Long Papers)}, 2018, pp. 323--333.

\bibitem{johnson2017clevr}
J.~Johnson, B.~Hariharan, L.~Van Der~Maaten, L.~Fei-Fei, C.~Lawrence~Zitnick, and R.~Girshick, ``Clevr: A diagnostic dataset for compositional language and elementary visual reasoning,'' in \emph{Proceedings of the IEEE conference on computer vision and pattern recognition}, 2017, pp. 2901--2910.

\bibitem{du2024paged}
W.~Du, W.~Liao, H.~Liang, and W.~Lei, ``{PAGED}: A benchmark for procedural graphs extraction from documents,'' in \emph{Proceedings of the 62nd Annual Meeting of the Association for Computational Linguistics (Volume 1: Long Papers)}, 2024, pp. 10\,829--10\,846.

\bibitem{miller1995wordnet}
G.~A. Miller, ``{WordNet:} a lexical database for english,'' \emph{Communications of the ACM}, vol.~38, no.~11, pp. 39--41, 1995.

\bibitem{gpt-4o-mini}
OpenAI, ``{GPT-4o} mini: advancing cost-efficient intelligence,'' \url{https://openai.com/index/gpt-4o-mini-advancing-cost-efficient-intelligence/}.

\bibitem{hurst2024gpt}
A.~Hurst, A.~Lerer, A.~P. Goucher, A.~Perelman, A.~Ramesh, A.~Clark, A.~Ostrow, A.~Welihinda, A.~Hayes, A.~Radford \emph{et~al.}, ``{GPT-4o} system card,'' \emph{arXiv preprint arXiv:2410.21276}, 2024.

\bibitem{li2022competition}
Y.~Li, D.~Choi, J.~Chung, N.~Kushman, J.~Schrittwieser, R.~Leblond, T.~Eccles, J.~Keeling, F.~Gimeno, A.~Dal~Lago \emph{et~al.}, ``Competition-level code generation with {AlphaCode},'' \emph{Science}, vol. 378, no. 6624, pp. 1092--1097, 2022.

\bibitem{shi2022natural}
F.~Shi, D.~Fried, M.~Ghazvininejad, L.~Zettlemoyer, and S.~I. Wang, ``Natural language to code translation with execution,'' in \emph{Proceedings of the 2022 Conference on Empirical Methods in Natural Language Processing}, 2022, pp. 3533--3546.

\bibitem{franti2023soft}
P.~Fr{\"a}nti and R.~Mariescu-Istodor, ``Soft precision and recall,'' \emph{Pattern Recognition Letters}, vol. 167, pp. 115--121, 2023.

\bibitem{reimers2019sentence}
\BIBentryALTinterwordspacing
N.~Reimers and I.~Gurevych, ``{Sentence-BERT}: Sentence embeddings using siamese {BERT-Networks},'' in \emph{Proceedings of the 2019 Conference on Empirical Methods in Natural Language Processing}.\hskip 1em plus 0.5em minus 0.4em\relax Association for Computational Linguistics, 11 2019. [Online]. Available: \url{http://arxiv.org/abs/1908.10084}
\BIBentrySTDinterwordspacing

\bibitem{artifact_models2025}
\BIBentryALTinterwordspacing
B.~Chen, O.~Wei, B.~Zheng, and G.~Mussbacher, ``Abscon paperartifacts.'' [Online]. Available: \url{https://github.com/20001LastOrder/LLM-AbsCon}
\BIBentrySTDinterwordspacing

\bibitem{meissel2024using}
K.~Meissel and E.~S. Yao, ``Using {Cliff}’s delta as a non-parametric effect size measure: an accessible web app and r tutorial,'' \emph{Practical Assessment, Research, and Evaluation}, vol.~29, no.~1, 2024.

\bibitem{rabbani2023extraction}
K.~Rabbani, M.~Lissandrini, and K.~Hose, ``Extraction of validating shapes from very large knowledge graphs,'' \emph{Proceedings of the VLDB Endowment}, vol.~16, no.~5, pp. 1023--1032, 2023.

\bibitem{nair2024midgard}
I.~Nair and L.~Wang, ``{MIDGARD:} self-consistency using minimum description length for structured commonsense reasoning,'' in \emph{Proceedings of the 62nd Annual Meeting of the Association for Computational Linguistics, {ACL} 2024}.\hskip 1em plus 0.5em minus 0.4em\relax Association for Computational Linguistics, 2024, pp. 7047--7065.

\bibitem{bauer2013weighted}
S.~S. Bauer, U.~Fahrenberg, L.~Juhl, K.~G. Larsen, A.~Legay, and C.~Thrane, ``Weighted modal transition systems,'' \emph{Formal Methods in System Design}, vol.~42, pp. 193--220, 2013.

\bibitem{salay2013managing}
R.~Salay, M.~Chechik, J.~Horkoff, and A.~D. Sandro, ``Managing requirements uncertainty with partial models,'' \emph{Requirements Engineering}, vol.~18, pp. 107--128, 2013.

\bibitem{wan2024dynamic}
G.~Wan, Y.~Wu, J.~Chen, and S.~Li, ``Dynamic self-consistency: Leveraging reasoning paths for efficient {LLM} sampling,'' \emph{arXiv preprint arXiv:2408.17017}, 2024.

\bibitem{yang2024multi}
Y.~Yang, B.~Chen, K.~Chen, G.~Mussbacher, and D.~Varr{\'o}, ``Multi-step iterative automated domain modeling with large language models,'' in \emph{Proceedings of the ACM/IEEE 27th International Conference on Model Driven Engineering Languages and Systems Companion (MODELS-C)}, 2024, pp. 587--595.

\bibitem{jahan2024automated}
M.~Jahan, M.~M. Hassan, R.~Golpayegani, G.~Ranjbaran, C.~Roy, B.~Roy, and K.~Schneider, ``Automated derivation of {UML} sequence diagrams from user stories: Unleashing the power of generative {AI} vs. a rule-based approach,'' in \emph{Proceedings of the ACM/IEEE 27th International Conference on Model Driven Engineering Languages and Systems}, 2024, pp. 138--148.

\bibitem{sultan2024ai}
B.~Sultan and L.~Apvrille, ``{AI}-driven consistency of {SysML} diagrams,'' in \emph{Proceedings of the ACM/IEEE 27th International Conference on Model Driven Engineering Languages and Systems}, 2024, pp. 149--159.

\bibitem{lopez2024text2vql}
J.~A.~H. L{\'o}pez, M.~F{\"o}ldi{\'a}k, and D.~Varr{\'o}, ``{Text2VQL}: Teaching a model query language to open-source language models with {ChatGPT},'' in \emph{Proceedings of the ACM/IEEE 27th International Conference on Model Driven Engineering Languages and Systems}, 2024, pp. 13--24.

\end{thebibliography}

\end{document}